\documentclass[conference]{IEEEtran}
\IEEEoverridecommandlockouts
\usepackage{cite}
\usepackage{amsmath,amssymb,amsfonts}
\usepackage{graphicx}
\usepackage{textcomp}
\usepackage{xcolor}
\usepackage{graphicx}
\usepackage{float}
\usepackage{subfigure}
\usepackage{caption}
\usepackage{multirow}
\usepackage{longtable}
\usepackage{tabu}
\usepackage{colortbl,booktabs}
\usepackage{algorithm}
\usepackage{algorithmicx}
\usepackage[noend]{algpseudocode}
\usepackage{amsfonts,dsfont}
\usepackage{verbatim}
\usepackage{tabulary}
\usepackage{url}
\usepackage{fancyhdr}

\def\BibTeX{{\rm B\kern-.05em{\sc i\kern-.025em b}\kern-.08em
    T\kern-.1667em\lower.7ex\hbox{E}\kern-.125emX}}


\errorcontextlines\maxdimen
\makeatletter
    \newcommand*{\algrule}[1][\algorithmicindent]{\makebox[#1][l]{\hspace*{.5em}\thealgruleextra\vrule height \thealgruleheight depth \thealgruledepth}}%
\newcommand*{\thealgruleextra}{}
\newcommand*{\thealgruleheight}{.75\baselineskip}
\newcommand*{\thealgruledepth}{.25\baselineskip}
\newcount\ALG@printindent@tempcnta
\def\ALG@printindent{%
    \ifnum \theALG@nested>0
        \ifx\ALG@text\ALG@x@notext
        \else
            \unskip
            \addvspace{-1pt}
            \ALG@printindent@tempcnta=1
            \loop
                \algrule[\csname ALG@ind@\the\ALG@printindent@tempcnta\endcsname]%
                \advance \ALG@printindent@tempcnta 1
            \ifnum \ALG@printindent@tempcnta<\numexpr\theALG@nested+1\relax
            \repeat
        \fi
    \fi
    }%
\usepackage{etoolbox}
\patchcmd{\ALG@doentity}{\noindent\hskip\ALG@tlm}{\ALG@printindent}{}{\errmessage{failed to patch}}
\makeatother
\newbox\statebox
\newcommand{\myState}[1]{%
    \setbox\statebox=\vbox{#1}%
    \edef\thealgruleheight{\dimexpr \the\ht\statebox+1pt\relax}%
    \edef\thealgruledepth{\dimexpr \the\dp\statebox+1pt\relax}%
    \ifdim\thealgruleheight<.75\baselineskip
        \def\thealgruleheight{\dimexpr .75\baselineskip+1pt\relax}%
    \fi
    \ifdim\thealgruledepth<.25\baselineskip
        \def\thealgruledepth{\dimexpr .25\baselineskip+1pt\relax}%
    \fi
    \State #1%
    \def\thealgruleheight{\dimexpr .75\baselineskip+1pt\relax}%
    \def\thealgruledepth{\dimexpr .25\baselineskip+1pt\relax}%
}

\newtheorem{definition}{Definition}

\begin{document}

\title{EnPAC: Petri Net Model Checking for Linear Temporal Logic\\
\thanks{This work is partially supported by National Key Research and Development Program of China under Grant No.2022YFB4501700 and National Natural Science Foundation of China under Grant No.61672381.}
}

\author{\IEEEauthorblockN{Zhijun Ding, Cong He, and Shuo Li}
\IEEEauthorblockA{\textit{Department of Computer Science and Technology}\\ 
\textit{Tongji University}\\ Shanghai, China\\ 
dingzj@tongji.edu.cn, 1105585684@qq.com, and lishuo20062005@126.com
}}

\maketitle

\thispagestyle{plain}

\begin{abstract}
State generation and exploration (counterexample search) are two cores of explicit-state Petri net model checking for linear temporal logic (LTL). Traditional state generation updates a structure to reduce the computation of all transitions and frequently encodes/decodes to read each encoded state. We present the optimized calculation of enabled transitions on demand by dynamic fireset to avoid such a structure. 
And we propose direct read/write (DRW) operation on encoded markings without decoding and re-encoding to make state generation faster and reduce memory consumption. 
To search counterexamples more quickly under an on-the-fly framework, we add heuristic information to the Büchi automaton to guide the exploration in the direction of accepted states. 
The above strategies can optimize existing methods for LTL model checking. 
We implement these optimization strategies in a Petri net model-checking tool called EnPAC (Enhanced Petri-net Analyser and Checker) for linear temporal logic. Then, we evaluate it on the benchmarks of MCC (Model Checking Contest), which shows a drastic improvement over the existing methods. 
\end{abstract}


\section{Introduction}\label{sec_introduction}

Model checking is a highly automatic technology based on a formalism, like Petri nets \cite{Murata1989Petri}, for verifying finite-state concurrent systems \cite{Wolf2019How}. 
Actually, many important temporal characteristics or functional requirements of concurrent systems are specified by linear temporal logics (LTLs) \cite{Gastin2001Fast}. 
In traditional LTL model checking, a formal system model is synchronized by using the product construction with Büchi automaton \cite{Gastin2001Fast} representing all behaviors that violate an LTL formula. Then, the existence of a run with infinitely many occurrences of an accepting state in the product automaton provides a counterexample for the LTL formula \cite{Vardi1996An} with an on-the-fly framework \cite{Geldenhuys2005More}. 

However, the state-explosion problem \cite{Valmari1992A} is the main obstacle to practical model checking, as the number of reachable states is exponentially larger than the size of a system description via Petri nets. Even the complexity of LTL model checking is exponential. So far, a lot of reduction techniques, e.g., abstraction, partial order reduction, and symmetry reduction, can decrease the size of the state space. 
Also, many mature tools (e.g., LoLA \cite{Liebke2020Faster}) of Petri net model checking implement efficient state generation and exploration techniques. 
We focus on optimizing state generation and exploration strategies in existing tools in this paper. 

Since model checking is essentially an exhaustive exploration technology on state space, state generation is the core of the whole process. 
Traditional methods must calculate and store all enabled transitions under each reachable state. However, many enabled transitions may never occur under on-the-fly exploration. Thus, computing all transitions and storing all enabled transitions lead to a waste of time and memory. Addressing this problem, LoLA designs a data structure \cite{Liebke2020Faster} to accelerate computing all enabled transitions $T(m)$ by briefly updating $T(m')$ when migrating from a state $m'$ to the next state $m$. Although it can avoid computing all transitions by such a static structure, it brings some memory cost. 
In this paper, we optimize the calculation of enabled transitions under each state. It is more efficient to calculate the enabled transitions on demand. 
We propose the first optimized strategy of dynamically calculating the transition set, where only one enabled transition is calculated when a successor state is generated. 

Most explicit-state Petri net model-checking tools exploit various encoding strategies in marking storage, saving large memory costs. Then, when calculating the enabled transitions, they require reading the number of tokens of particular places in the encoded marking. 
To our knowledge, they should have a decoding and encoding procedure when reading or writing encoded markings as shown in, e.g., LoLA \cite{Wolf2018Petri}. However, frequent decoding and re-encoding based on an encoding strategy can reduce tool efficiency. It is challenging to read the number of tokens directly by reading and writing encoded markings. 
We define a reading pattern and a writing pattern for each place and propose a set of bitwise operations on our new pattern and the encoded markings. Thus, we propose the second optimized strategy of direct read/write (DRW) operations. 

State exploration (counterexample search) is another core for explicit LTL model checking. 
Actually, the faster the counterexample is found, the fewer states are generated with an on-the-fly framework. To find a counterexample faster than random exploration, it is better to reach an acceptable state faster. Based on this insight, we present the third optimized strategy to add a heuristic to the Büchi automaton. The heuristic can guide the on-the-fly exploration in the direction of accepted states. 

Based on the above optimization insights, we implement an explicit-state model-checking tool EnPAC standing for Enhanced Petri-net Analyser and Checker. It can be used for large concurrent systems, modeled as Petri nets \cite{Reisig2012Petri} or  Colored Petri Nets (CPNs) as its colored extension\cite{Jensen2013Coloured}. It can evaluate arbitrary queries specified in linear temporal logic. 
Then, we evaluate the performance on the benchmarks in Model Checking Contest (MCC) \cite{MCC}. 
These optimized state generation and exploration strategies help EnPAC make excellent  progress and drastic improvement on the benchmarks of MCC \cite{MCC}. 
The contributions are three optimization strategies summarized as follows.

1. We propose a dynamic fireset to calculate enabled transitions on demand, avoiding traditional complete enabled transition calculation and additional data structure required by LoLA \cite{Liebke2020Faster}. 

2. We propose a new reading and writing pattern for each place by a set of bitwise operations (called DRW operations) on each marking without frequent decoding and encoding on the encoded marking storage.

3. We present a heuristic Büchi automaton to guide the exploration for searching a counterexample faster, which helps avoid traditional random exploration.

We introduce the preliminaries in Section \ref{sec_preliminary} and detail the proposed optimizations in Section \ref{sec_optimizations}. 
Our experiment results are evaluated based on EnPAC in Section \ref{sec_experiment}. 
Finally, this paper is concluded in Section \ref{sec_conclusion}. 


\section{Preliminary}\label{sec_preliminary}
\subsection{Petri Nets}
Petri nets have been widely used in the modeling and verification of concurrent systems for many interesting properties of concurrent systems, such as deadlock, liveness, and reachability. 
We first introduce the definition of Petri net.

\begin{definition}\label{def_pn}
A Petri net $N$ is a five-tuple $N=\{P,T,F,W,m_0\}$ where $P$ is a finite set of places, $T$ is a finite set of transitions (disjoint to $P$), $F\subseteq\left(P\times T\right)\cup(T\times P)$ is a finite set of arcs, $W\colon\left(P\times T\right)\cup\left(T\times P\right)\rightarrow\mathbb{N}$ is a weight function where $(x,y)\notin F \iff W\left(x,y\right)=0$, and $m_0$ is the initial marking. A marking is a mapping $m:P\rightarrow\mathbb{N}$.
\end{definition}

\begin{definition}\label{def_behavior_pn}
A transition $t$ is enabled under a marking $m$ if $\forall p\in P$,$\ W\left(p,t\right)\le m(p)$. We call the set of all enabled transitions in $m$ \textit{fireset}, denoted by $T(m)$. Firing an enabled transition $t$ under a marking $m$ leads to a new marking $m'$ where $m'\left(p\right)=m\left(p\right)-W\left(p,t\right)+W(t,p)$. This firing relation is denoted as $m\xrightarrow{t}m'$. If there exists a transition sequence $\omega=t_1t_2\cdots t_n$ such that $m_1\xrightarrow{t_1}m_2\xrightarrow{t_2}\cdots\xrightarrow{t_n}m_n$, $m_n$ is reachable from $m_1$, written $m_1\xrightarrow{*}m_n$. The state space of a Petri net consists of $R\left(m_0\right)=\{m\mid m_0\xrightarrow{*}m\}$. 
\end{definition}


\subsection{Linear Temporal Logic}

We define the syntax and semantics of atomic proposition based on MCC \cite{MCC}, and then LTL. 

\begin{definition}\label{def_AP}
Let $\langle atomic\rangle$ be an atomic proposition, and $\langle int\mbox{-}expression\rangle$ be an expression evaluated by an integer. 

\begin{align*}\centering
\langle atomic\rangle&:=is\mbox{-}fireable(t_1,\cdots,t_n) \\
& \vert\langle int\mbox{-}expression\rangle \\
& \leq\langle int\mbox{-}expression\rangle \\
\langle int\mbox{-}expression\rangle &:=Int\vert tokens\mbox{-}count(p_1,\cdots,p_n)
\end{align*}

$is\mbox{-}fireable\left(t_1,\cdots,t_n\right)$ holds if either $t_1$ or $t_2$ or $\cdots$ or $t_n$ are enabled, and $tokens\mbox{-}count(p_1,\cdots,p_n)$ returns the exact number of tokens contained in the place set $\{p_1,\cdots,p_n\}$.
\end{definition}

\begin{definition}\label{def_ltl}
Every atomic proposition is an LTL formula. If $\varphi$ and $\psi$ are LTL formulae, so are $\lnot\varphi$, $\ \left(\varphi\vee\psi\right)$, $\left(\varphi\land\psi\right)$, $X\varphi$, $F\varphi$, $G\varphi$, $\left(\varphi U\psi\right)$, $\left(\varphi R\psi\right)$. Let $AP$ be a non-empty finite set of atomic propositions, $\xi=x_0x_1x_2\cdots$ be a sequence over alphabet $2^{AP}$, $\varphi$ and $\psi$ be LTL formulae. We write $\xi_i$ for the suffix of $\xi$ starting at $x_i$. $\xi$ satisfies an LTL formula according to the following inductive scheme: $\xi\models p\iff p\in x_0,\ p\in AP$; $\xi\models\lnot\varphi\iff \xi$ dissatisfy $\varphi$; $\xi\models\varphi\vee\psi\iff \xi\models\varphi\ \mathrm{or} \ \models\psi$; $\xi\models X\varphi\iff \xi_1\models\varphi$; $\xi\models\varphi U\psi\iff i\geq0,\ \xi_i\models\psi\land(\forall j<i,\ \xi_i\models\varphi)$;
Other operators $(\wedge,R,F,G)$ can be derived from the above operators: $\varphi\land\psi\equiv\lnot(\lnot\varphi\vee\lnot\psi)$; $\varphi R\psi\equiv\lnot(\lnot\varphi U\lnot\psi)$; $F\varphi\equiv\left(TRUE\right)U\varphi$; $G\varphi\equiv\lnot(F\lnot\varphi\ )$
\end{definition}




\subsection{On-the-fly Exploration of LTL Model Checking}\label{sec_architechture}

\begin{figure}[h]\centering
\includegraphics[width=0.48\textwidth]{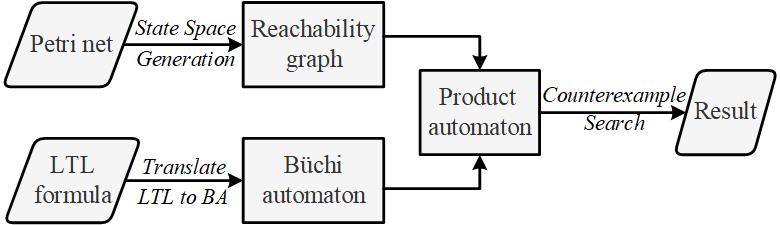}\caption{The entire process of explicit-state LTL model checking}\label{fig_flowchart}
\end{figure}

As shown in Fig. \ref{fig_flowchart}, explicit-state LTL model-check tools for Petri nets calculate the reachability graph of a Petri net, transform the negative LTL formula into the Büchi automaton, and then generate the product automaton in the form of Cartesian product of reachability graph and Büchi automaton, and finally searches counterexamples on the product automaton. 
If a counterexample is explored, $false$ is returned. Otherwise, $true$ is returned.

This process usually uses an on-the-fly framework to optimize the above process. On-the-fly exploration \cite{Geldenhuys2005More} consists in constructing a reachability graph and product automaton while checking for the counterexamples in the product automaton. An advantage of on-the-fly exploration is that it can return a result before the entire state space is constructed.

\section{Optimization Strategies}\label{sec_optimizations}

Since explicit model checking is an exploration technology on state space, the efficiency of state generation and exploration (counterexample search) directly affects its performance.  
Clearly, improving the efficiency of state generation and exploration is vitally important. 
We propose two optimizations to make state generation much faster with less memory consumption. One is dynamic calculating enabled transitions. The other is direct reading and writing encoded markings without frequent decoding and re-encoding procedures. 
Concerning state exploration, we propose a heuristic Büchi automaton as an optimization to guide on-the-fly exploration to explore counterexamples in the direction of accepted states. 
\subsection{Dynamic Fireset}\label{subsec_dynamic_fireset}
Enabled transitions play a fundamental role in the state generation since they determine all successor states of a reachable state. The common practice is generating all enabled transitions simultaneously and firing one by one to enumerate all possible successor states. 
We generate only one enabled transition at a time in a state. Then, when on-the-fly exploration backtracks to an explored state, how to directly generate the next enabled transition without repeated exploration is a difficulty.

To solve this difficulty, we fix it by defining a total order $(\prec, T)$ on the transition set and then checking which transition is enabled in turn in that order. We also define an array to store all the transitions and use their index as their total order. 
In addition, each explored state needs to record the last fired transition. In this case, when on-the-fly exploration backtracks to an explored state, it can check the transition just next to the last fired transition by order $(\prec, T)$ to fire the next enabled transition. 
Once an enabled transition $t$ is found, on-the-fly exploration stops to generate a successor state $m$ by firing $t$ and continues a depth-first search on $m$. 
Our new method is named \textit{dynamic fireset} (abbreviated as DYN). Its advantage is that it saves memory and time to calculate and store all enabled transitions under each reachable state. 


\subsection{DRW Operation on Encoded Markings}\label{subsec_direct_read/write}

\begin{figure}[h]\centering
\includegraphics[width=0.45\textwidth]{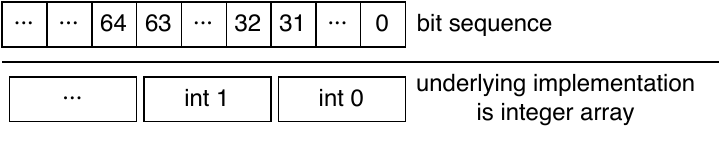}
\caption{Underlying implementation of bit sequence} \label{fig_bitsequence}
\end{figure}

Our encoding strategy is designed on an integer array, as shown in Fig. \ref{fig_bitsequence}, which is an underlying implementation of a bit sequence. 
Based on it, we propose a new method of DRW operation on the encoded marking, which can be divided into two sub-tasks. One is to locate, i.e., in which integer the place's coding resides and from which bit of that integer it begins. The other is to read/write its value. 
To locate the correct position, we record each place's start position in the bit sequence and its length. To read or write a correct value, we define a reading pattern and a writing pattern for each place. Then, the token counts can be easily read or written by a series of bitwise operations using these patterns. 

There are four kinds of marking encoding in APPENDICES \ref{app_pn}.
We use NUPN \cite{Garavel2015Nested} encoding to illustrate our DRW operations.
In it, each place carries two extra attributes, $myunit$, and $myoffset$, indicating the unit in which it is located and the offset number in the unit, respectively. Each unit carries two attributes, too, $startpos$ and $unitlen$, indicating the start position in the bit sequence and how many bits this unit takes. When reading or writing a place, there are two cases. In other words, the encoding of the unit occupies only one integer or spans two integers. The algorithms for the two cases are detailed in APPENDICES \ref{app_core_modules}.

\subsection{Heuristic Büchi Automaton}\label{subsec_heuristicBA}

\begin{figure}[t]\centering
\includegraphics[width=0.38\textwidth]{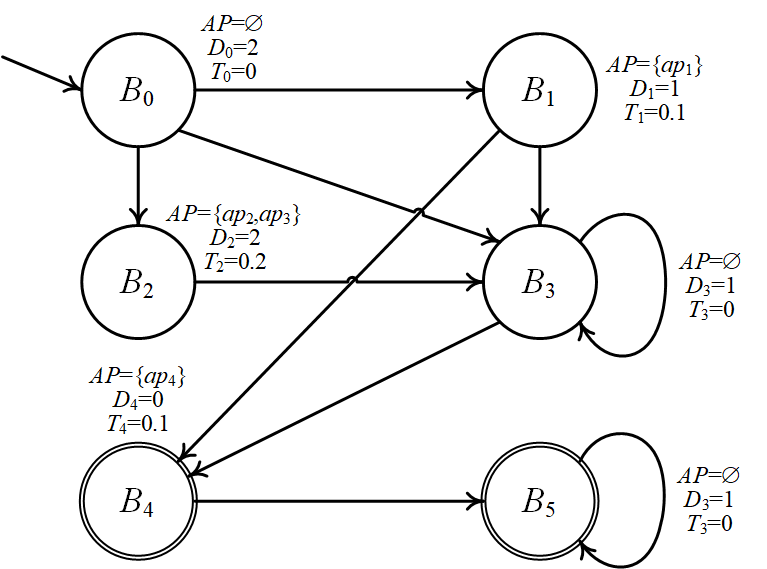}\caption{Büchi automaton with heuristic information}\label{fig_heuristicBA}
\end{figure}

\begin{figure*}[t]\centering
\includegraphics[width=0.93\textwidth]{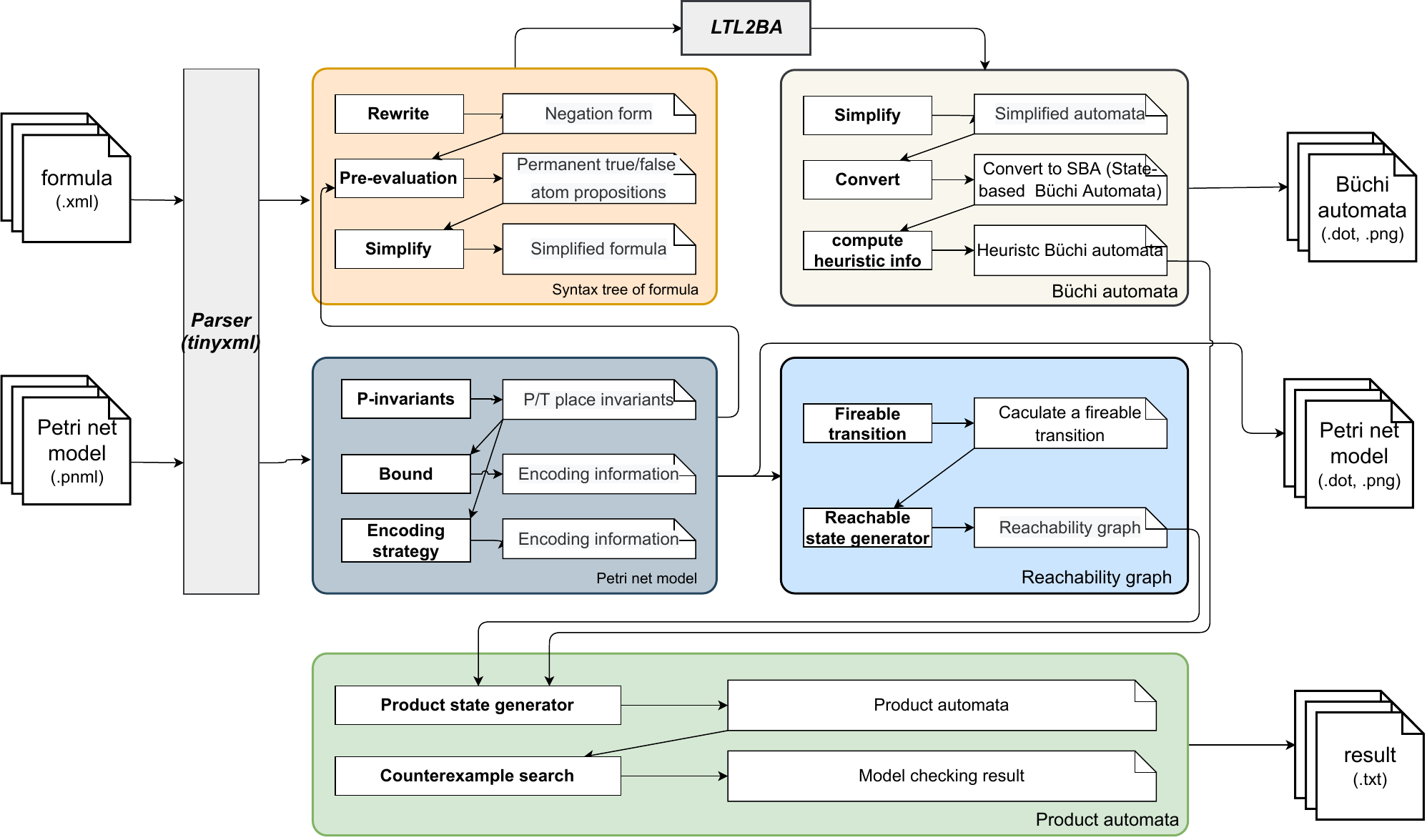}\caption{The architechture of EnPAC}\label{fig_architechture}
\end{figure*}

Before state exploration, the Büchi automaton is automatically generated, which has complete information. We propose the heuristic in Büchi automaton in two aspects. 
Firstly, searching counterexamples in a product automaton is to find a strongly connected component containing accepted states. And whether a product state $(m_i,b_i)$ is accepted is determined by its Büchi state part $b_i$. When generating a reachable state and seeking a Büchi state to combine, it should choose the state that reaches an accepting state the fastest. 
Thus, the distance to an accepting state is the first aspect we consider.
As for each state in the Büchi automaton, the number of atomic propositions affects how easy to synthesize this state.

Based on these insights, we add two extra attributes $D_i$ and $T_i$, as the heuristics in each Büchi state. 
Concretely, $D_i$ is the length of the shortest path from state $B_i$ to an acceptable state. Take Fig. \ref{fig_heuristicBA} as an example, $D_0$ in $B_0$ is 2 because its shortest path to an accepted state is $B_0 \rightarrow B_1 \rightarrow B_4$ (or $B_0 \rightarrow B_3 \rightarrow B_4$) whose length is $2$. 
And $D_i$ can be computed by Dijkstra's algorithm.
Let $AP_i$ be the set of atomic propositions carried by Büchi state $B_i$, and $T_i$ $|AP_i|*0.1$ ($T_i$ is the number of atomic propositions carried by $B_i$. The coefficient '0.1' is from our experience on MCC Benchmark).
$T_i$ indicates how tough it is for state $B_i$ to produce a reachable state into a product state. 

In our heuristic Büchi automaton (abbreviated as HBA), when choosing a Büchi state for the product with a reachable state, we prioritize the state with smaller $D_i+T_i$. It means we always prefer the path that can reach an accepted state fast and be smooth enough. 

\section{Experimental Evaluation}\label{sec_experiment}

\subsection{Installation and Usage}\label{subsec_installation}

We implement our optimizations in EnPAC (Enhanced Petri-net Analyser and Checker).
It is divided into five modules, including the Petri net model, Reachability graph, Syntax tree of LTL formula, Büchi automaton, and Product automaton. The architecture of EnPAC is shown in Fig. \ref{fig_architechture}, detailed in APPENDICES \ref{app_core_modules}.

EnPAC can be downloaded from \url{https://github.com/Tj-Cong/EnPAC_2021} and installed easily. The GitHub homepage presents a user manual that describes the installation procedure, file formats, output, and options. 
EnPAC can be utilized on the command line of the Linux terminal. The results can be displayed on the screen or in a file.  

\subsection{Benchmarks and Methodology}\label{subsec_bench}

For evaluating the optimization strategies via EnPAC, we use the benchmarks provided by MCC \cite{MCC}. 
The benchmarks consist of $1016$ Petri net instances, as well as $32$ LTL formulae per instance ($32512$ LTL formulae). 

In the benchmarks, there are $3672$ LTL formulae that no tool could give a result in MCC'2020 \cite{mcc:2020}. 
EnPAC has not yet implemented \textit{dynamic fireset} (abbreviated as DYN), \textit{direct read/write operation} (abbreviated as DRW), and \textit{heuristic Büchi automaton} (abbreviated as HBA) for MCC'2020 \cite{mcc:2020}. 
Thus, we compare our three optimizations with the results in MCC'2020 \cite{mcc:2020} to ensure that the results are persuasive. 
We call the version without the implementation of our optimizations the \textit{original} method (abbreviated as ORI).


\subsection{Experimental Analysis}\label{subsec_experiments}
For verifying each formula, there is a time limit of $300$ seconds and a memory limit of $16GB$. We use each optimization individually to illustrate their performances. The experimental results are shown in APPENDICES \ref{app_results}.

\subsubsection{Experiments for Dynamic Fireset}\label{subsub_DF}

In order to show the effect of DYN clearly, the time and memory peak for each formula are recorded.  
TABLE \ref{tab_DYN} shows the comparison results between the dynamic fireset method (DYN) and the original method (ORI) on $8$ Petri nets instances (their names are in the first column). Two LTL formulae are verified for each instance in the second column. 
Concretely, $T_{ORI}$ and $T_{DYN}$ are the whole time of ORI and DYN, respectively. And $M_{ORI}$ and $M_{DYN}$ are the memory peak of ORI and DYN, respectively. 
To quantify the optimized performance for DYN, we calculate $\nabla T_1$ by $T_{ORI}/T_{DYN}$, and $\nabla M_1$ by $M_{ORI}/M_{DYN}$ in TABLE \ref{tab_DYN}. The average of each result is shown in the last row. 
All results come to the same conclusion that our DYN outperforms ORI on time and memory consumption. 

It can be found from the experimental results that our DYN method is slightly faster than the ORI method in most instances. 
In particular, an LTL formula of 'CircadianClock-PT-001000' is originally timed out, but our optimization of DYN can output the result within $130$s. Except for this result of a timeout, DYN has an average improvement on time of $3.41$ times. 
Moreover, because DYN does not need to store all enabled transitions in every reachable state, it uses much less memory consumption than ORI. 

\subsubsection{Experiments for DRW Operations}\label{subssub_DRWO}

Due to different DRW operations for our encoding strategies as explained in APPENDICES \ref{app_pn}, we conduct separate experiments on 1-safe encoding ($8$ instances), NUPN encoding ($10$ instances), and P-invariant encoding ($5$ instances) in the first column. For each instance in the second column, there are also two LTL formulae in the third column. $T_{ORI}$ and $T_{DRW}$ are the whole time of ORI and DRW, respectively. And $M_{ORI}$ and $M_{DRW}$ are the memory peak of ORI and DRW, respectively. 
To quantify the optimized performance for DRW, we calculate $\nabla T_2$ by $T_{ORI}/T_{DRW}$ and $\nabla M_2$ by $M_{ORI}/M_{DRW}$ in TABLE \ref{tab_DRW}. The average of each result is shown in the last row.

It can be found that DRW is much faster than ORI on time. 
For 1-safe encoding, DRW outperforms ORI by more than $20$ times in $9$ formulae. 
Especially for NUPN encoding, DRW outperforms ORI in all formulae on time. 
In P-invariant encoding, DRW outperforms ORI by more than $200$ times in $4$ formulae. 
And DRW has an average improvement of $245.52$ times than ORI.
However, our DRW method uses slightly more memory because it requires additional space overhead for the read/write patterns of each place. 
But such costs are minuscule since the average of $\nabla M_2$ is mostly close to $1$.

\subsubsection{Experiments for Heuristic Büchi Automaton}\label{subsub_Heu}

In addition to time and memory, we add a comparison of the reachable states that need to be generated to find counterexamples. It can reflect whether the heuristic Büchi automaton can guide on-the-fly exploration to find the counterexample faster. 
TABLE \ref{tab_HBA} shows the experimental results on $10$ instances in the first column with two verified LTL formulae in the second column. Concretely, $N_{ORI}$ and $N_{HBA}$ are the state counts, $T_{ORI}$ and $T_{HBA}$ are the whole time, and $M_{ORI}$ and $M_{HBA}$ are the memory peak of ORI and HBA, respectively. 
We also calculate $\nabla N$ by $N_{ORI}/T_{HBA}$, $\nabla T_3$ by $T_{ORI}/T_{HBA}$, and $\nabla M_3$ by $M_{ORI}/M_{HBA}$. 

In TABLE \ref{tab_HBA}, there are $4$ formulae that are originally timed out. And HBA outputs the results successfully with few states. Obviously, our heuristic Büchi automaton helps find counterexamples faster.
Due to generating fewer states, they also consume less memory. 
The average of $\nabla T_3$ is $6.4$. 
Although the heuristic information does not lead well to finding the counterexample for many other formulae, it does not produce large excessive costs in time and memory. Most of $\nabla M_3$ are $1.00$.

\subsection{Discussion}\label{subsec_discussion}
\begin{figure*}[t]\centering
\subfigure[DRW vs. ORI on time]{
\includegraphics[width=0.49\textwidth]{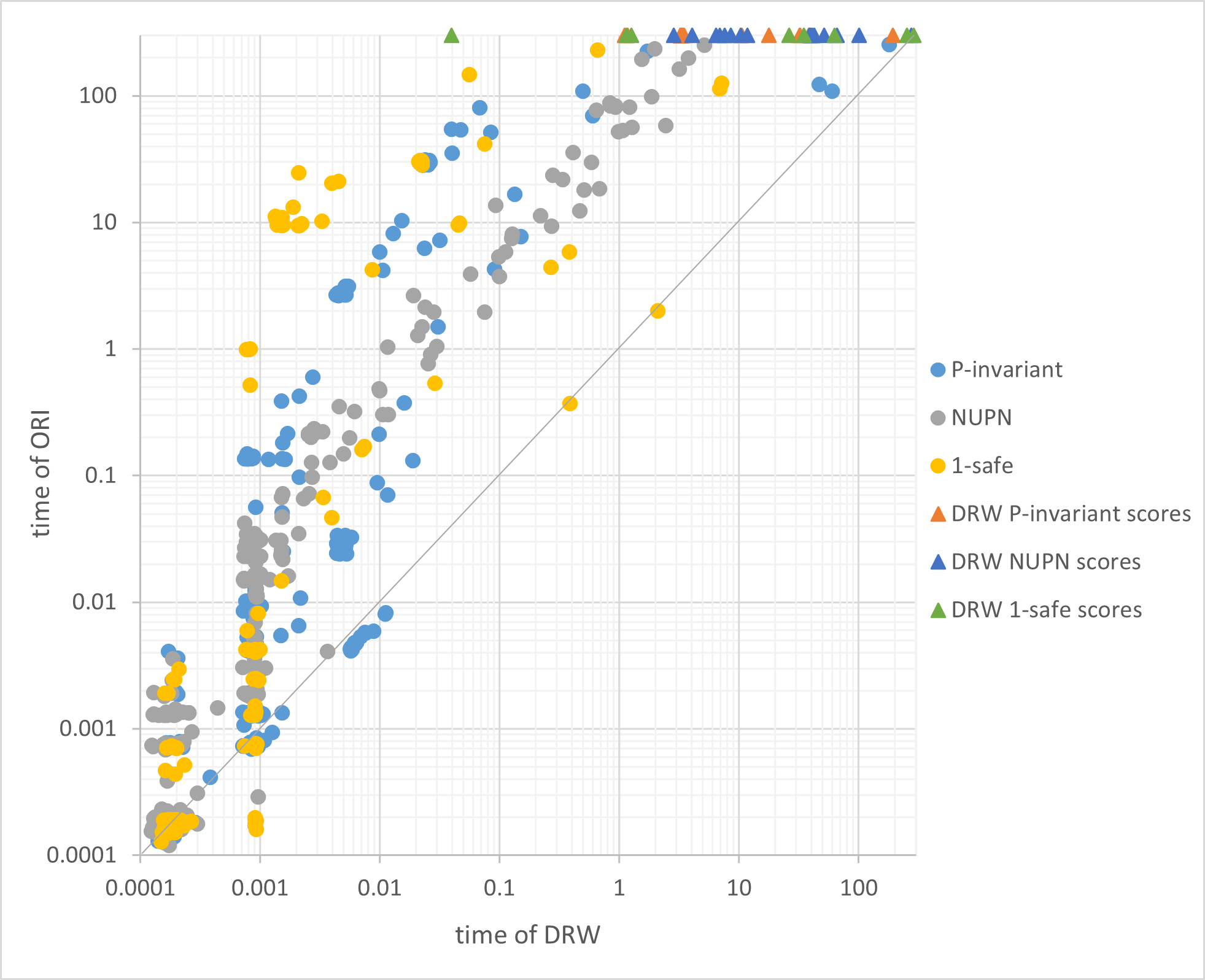}}
\subfigure[DRW vs. ORI on memory]{
\includegraphics[width=0.49\textwidth]{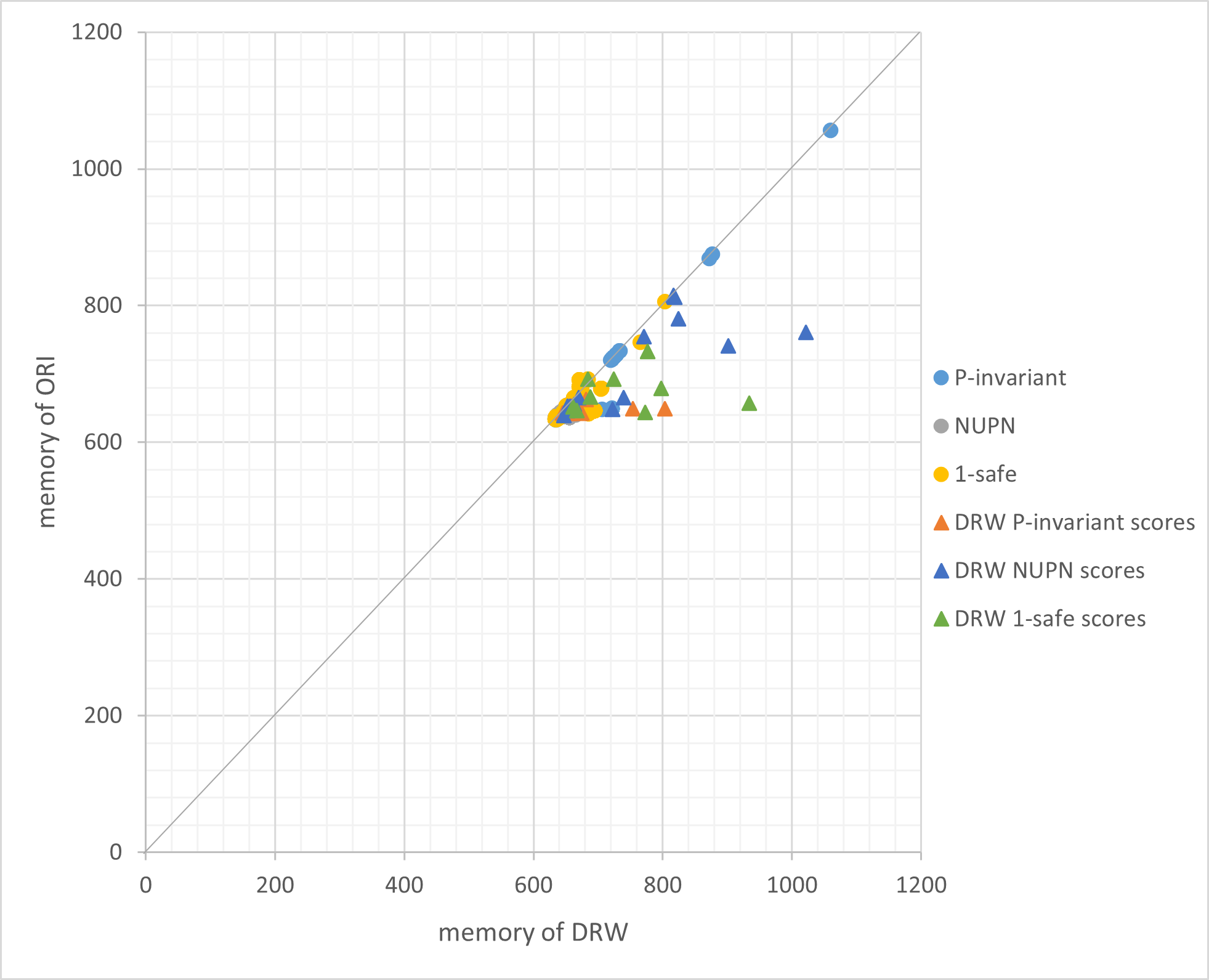}}
\\
\subfigure[DYN/HBA vs. ORI on time]{
\includegraphics[width=0.49\textwidth]{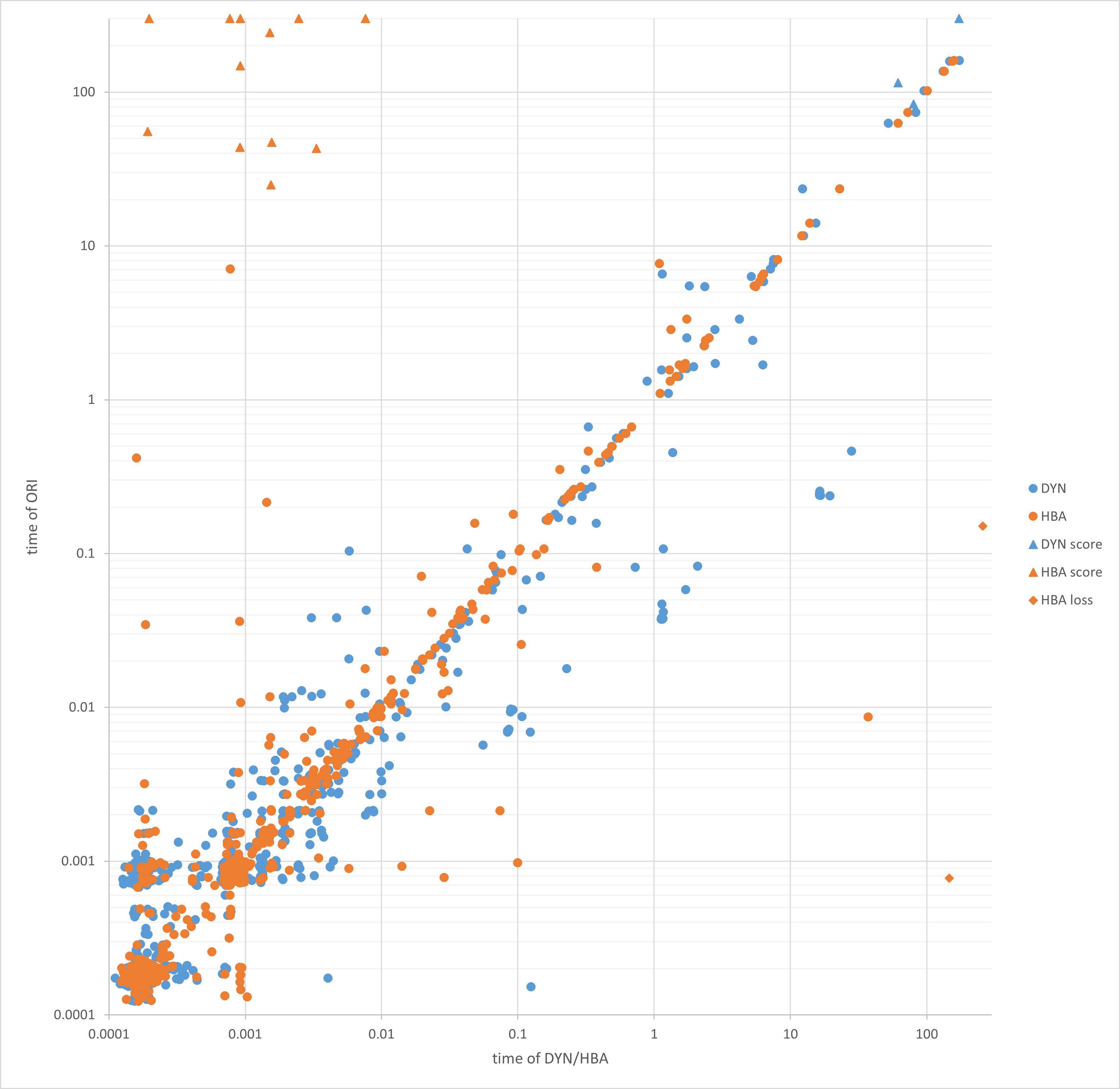}}
\subfigure[DYN/HBA vs. ORI on memory]{
\includegraphics[width=0.49\textwidth]{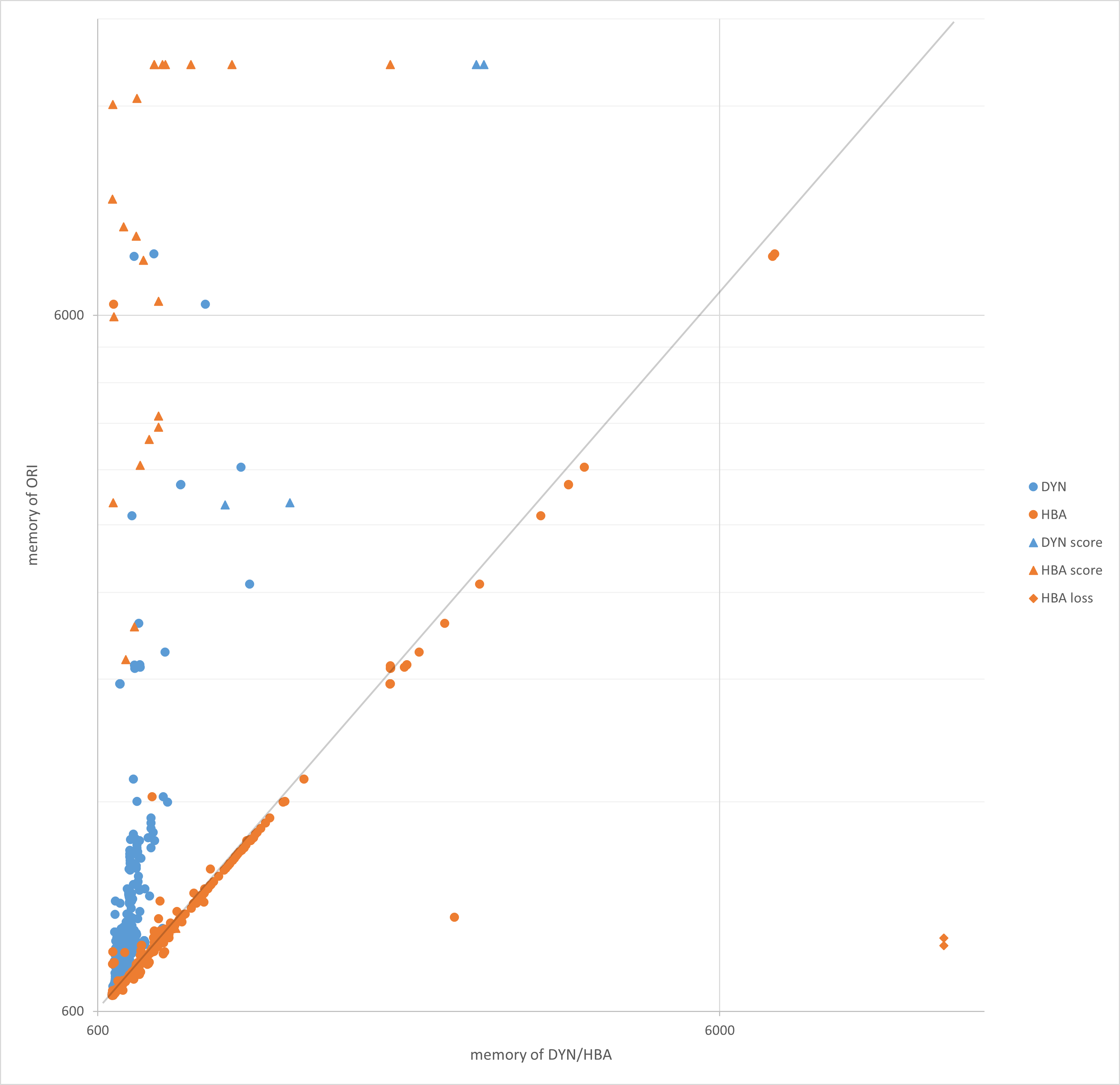}}
\caption{Comparison of original method and our optimization strategies}\label{fig_com}
\end{figure*}

We sketch four scatter plots in Fig. \ref{fig_com} on the benchmarks of MCC \cite{MCC}. 
The $x$-axis denotes the time/memory of DRW, DYN, and HBA, while the $y$-axis denotes the time/memory of ORI.
In the scatter plot, each dot represents an LTL formula verification, and the dots above the diagonal lines are the winning cases of our optimization. 

In Fig. \ref{fig_com}(a) and (b), the time and memory of DRW are demonstrated on all encoding methods, where 'P-invariant' represents P-invariant encoding, 'NUPN' represents NUPN encoding, and '1-safe' represents 1-safe encoding. 
We can see that DRW can significantly reduce time on most LTL formulae.
And the scores for DRW are the formula counts that ORI cannot output the result within $300s$ (although they have less memory in Fig. \ref{fig_com}(b)). It can be found that DRW can output the results within $150s$.

In Fig. \ref{fig_com}(c) and (d), the scores for DYN and HBA are also the formula counts that ORI cannot output the result within $300s$. 
Obviously, HBA works extremely well on some formulae, as we can see that some orange triangles and circles are much higher above the diagonal lines. 
Most orange dots are distributed near the diagonal lines. It confirms HBA does not produce large excessive costs in time and memory. HBA can also be counterproductive on individual formulae because it does not lead to counterexamples. 
Although most blue dots are distributed near the diagonal lines in Fig. \ref{fig_com}(c), DYN is much more effective in optimizing memory based on Fig. \ref{fig_com}(d).

There are $3,672$ formulae that no tool can output the results in MCC'2020 \cite{mcc:2020}. After using three optimization strategies, EnPAC can give results for $432$ unknown formulae. 
Under the complete benchmarks in MCC'2020 \cite{mcc:2020}, there are $28781$ LTL formulae that EnPAC has given the correct result before. 
With three optimizations, EnPAC correctly gives the results for $31735$ LTL formulae on the same benchmarks. Thus, our optimization strategies improve EnPAC by nearly $3,000$ scores, which shows a drastic improvement in EnPAC.

\section{Conclusion}\label{sec_conclusion}
We propose a dynamic fireset (DYN), which saves the storage and time of computing some redundant enabled transitions. We propose a direct read/write (DRW) operation on encoded marking, which saves the large overhead of decoding. In terms of state exploration, we add the heuristic information to the Büchi automaton (HBA) to guide the search of counterexamples, which speeds up the exploration.
We implement a tool called EnPAC for verifying LTL. We then evaluate it on the benchmarks of MCC. 
In the future, we improve the performance of EnPAC with more reduction techniques. 

\bibliographystyle{IEEEtran} 
\bibliography{IEEEabrv,enpac.bib} 

\newpage

\appendices

\section{Core Modules of EnPAC}\label{app_core_modules}

We briefly introduce the five core modules. EnPAC requires two input files, a PNML file that describes a Petri net and an XML file that specify LTL formulae. We use a third-party XML parser \textit{TinyXML-2} \cite{TinyXML2} for parsing the two input files.


\subsection{Petri net model}\label{app_pn}

In EnPAC, a Petri net is an object that contains three arrays storing $P$, $T$, and other elements in Definition \ref{def_pn}. In order to alleviate the state explosion, EnPAC uses a bit sequence to encode a marking instead of an integer vector, as shown in Fig. \ref{fig_bitsequence}. EnPAC adopts different encoding strategies for different Petri nets. Our encoding techniques include

\begin{itemize}
\item \textbf{default}: 16 bits are used per place. If any place's token number is over 65535 in the initial marking, EnPAC will allocate 32 bits per place. If EnPAC detects a reachable marking where a place's token number is over 65535 during the model-checking process, It will terminate, indicating it cannot handle the model.
	
\item \textbf{1-safe encoding}: 1-safe net is a Petri net where all place capacities are equal to one, indicating that each place can contain at most one token. EnPAC uses one bit to encode per place.
	
\item \textbf{NUPN encoding}: NUPN (Nested Unit Petri Net) \cite{Garavel2015Nested} is a special type of Petri net. Let $N=\{P,T,F,W,M_0,U,u_0,\sqsubseteq,unit\} $ be a NUPN. According to unit-safe property \cite{Garavel2019Nested}, for $\forall u_i\in U$, EnPAC use $\lceil log_2{(|u_i|+1)}\rceil$ bits to encode unit $u_i$ where $|u_i|$ indicates the number of local places of $u_i$.
	
\item \textbf{P-invariants encoding}: A place invariant is a mapping that assigns a weight to each place such that all reachable markings get the same weighted token sum. Place invariants divide all places into significant places and redundant places. The token number of redundant places can be calculated from significant places in any reachable marking. Also, place invariants can be used for estimating the upper bounds of places. We implemented the method in \cite{Schmidt2003Using,Wolf2019How} of using place invariants to estimate the upper bounds of places and encode markings. Let $S$ be the set of significant places of a Petri net, and $b(p_i)$ indicates the upper bound of place $p_i$. For $\forall p\in S$, EnPAC use $\lceil log_2{(b(p_i)+1)}\rceil$ bits to encode place $p_i$.
\end{itemize}

The Algorithm \ref{alg_direct_read_case1} and \ref{alg_direct_read_case2} are proposed for DRW operations in Section \ref{subsec_direct_read/write}.

\begin{algorithm*}[h]
\caption{Direct read/write within an integer}\label{alg_direct_read_case1}
\begin{algorithmic}[1]
\Function{read}{$p_i$,$array$}
		\State $index\gets unit\left[p_i.myunit\right].startpos/32$
		\State $offset\gets unit\left[p_i.myunit\right].startpos\%32$
		\State $pattern_{read}\gets (2^{unit\left[p_i.myunit\right].unitlen}-1)<<offset$ \Comment{'$<<$' is left shift operator}
		\State $value\gets \left(array\left[index\right]\& pattern_{read}\right)>>offset$ \Comment{'$>>$' is right shift operator}
		\If{$p_i.myoffset=value$} 
		\State $p_i$ is marked
		\Else
		\State $p_i$ is not marked
		\EndIf
		\EndFunction
		
		\Function{write}{$p_i,array$}
		\State $pattern_{zero}\gets \sim((2^{unit\left[p_i.myunit\right].unitlen}-1)<<offset)$ \Comment{'$\sim$' is bitwise NOT operator}
		\State $pattern_{write}\gets p_i.myoffset<<offset$
		\State $array\left[index\right]\gets\left(array\left[index\right]\& pattern_{zero}\right) \mid pattern_{write}$ \Comment{'$\mid$' is bitwise OR operator}
		\EndFunction
	\end{algorithmic}
\end{algorithm*}

\begin{algorithm*}[h]
\caption{Direct read/write across two integers}\label{alg_direct_read_case2}
\begin{algorithmic}[1]
\Function{read}{$p_i,array$}
		\State $index\gets unit\left[p_i.myunit\right].startpos/32$
		\State $offset\gets unit\left[p_i.myunit\right].startpos\%32$
		\State $pattern_{read\_low}\gets(2^{unit\left[p_i.myunit\right].unitlen}-1)<<offset$
		\State $pattern_{read\_high}\gets(2^{unit\left[p_i.myunit\right].unitlen}-1)>>(32-offset)$
		\State $value\gets\left(array\left[index\right]\& pattern_{read\_low}\right)>>offset$
		\Statex \ \ \ \ \ \ \ \ \ \ \ \ $+(array\left[index+1\right]\&pattern_{read_high})<<(32-offset)$
		\If{$p_i.myoffset=value$} 
		\State $p_i$ is marked
		\Else
		\State $p_i$ is not marked
		\EndIf
		\EndFunction
		
		\Function{write}{$p_i,array$}
		\State $pattern_{zero\_low}\gets\sim((2^{unit\left[p_i.myunit\right].unitlen}-1)<<offset)$
		\State $pattern_{zero\_high}\gets\sim((2^{unit\left[p_i.myunit\right].unitlen}-1)>>(32-offset))$
		\State $pattern_{write\_low}\gets p_i.myoffset<<offset$
		\State $pattern_{write\_high}\gets p_i.myoffset>>(32-offset)$
		\State $array\left[index\right]\gets\left(array\left[index\right]\& pattern_{zero\_low}\right) \mid pattern_{write\_low}$
		\State $array\left[index+1\right]\gets\left(array\left[index+1\right]\& pattern_{zero\_high}\right) \mid pattern_{write\_high}$
		\EndFunction
	\end{algorithmic}
\end{algorithm*}

\subsection{Reachability graph}\label{app_rg} 
The generation of the reachability graph is integrated into the on-the-fly framework. The main function of this module is to provide an interface for on-the-fly to generate reachable states and store reachability graphs. The reachability graph is stored in a chained hash table. To generate a successor of a reachable state $M$, fireable transitions under $M$ need to be calculated. For efficiency, EnPAC calculates only one fireable transition for every time on-the-fly backtracks to $M$. We call this method \textit{dynamic fireset}. One advantage of \textit{dynamic fireset} is that it saves much memory by avoiding storing all fireable transitions in every reachable marking. We detail it in Section \ref{subsec_dynamic_fireset}.

\subsection{Syntax tree of formula}\label{app_syntax_tree} 

After obtaining the syntax tree, EnPAC converts the formula into a negation form and simplifies it using a set of rewriting rules presented in \cite{Schmidt2000Calculate} to reduce the number of temporal operators. 

\begin{table}[h]\centering
\caption{Atomic Proposition Evaluation rules}\label{tab_APevaluation}
\resizebox{0.48\textwidth}{!}{
\begin{tabular}{clc}
\toprule
\textbf{Atomic proposition} &
	\multicolumn{1}{c}{\textbf{Condition}} &
	\multicolumn{1}{c}{\textbf{Evaluation}} \\ \midrule
	$0\le$tokens$\mbox{-}$count$(p_1,p_2,\cdots, p_n)$ &
	/ &
	$true$ \\ 
	tokens$\mbox{-}$count$\left(p_1,p_2,\cdots,p_n\right)\le k$ &
	\begin{tabular}[l]{@{}l@{}}1. the net is 1-safe\\ 2. $k\geq n$\end{tabular} &
	$true$ \\ 
	tokens$\mbox{-}$count$\left(p_1,p_2,\cdots,p_n\right)\le k$ &
	\begin{tabular}[l]{@{}l@{}}1. the net is a NUPN \\ 2. $k\geq|\{unit(p_i)\mid1\le i\le n\}|$\end{tabular} &
	$true$ \\ 
	$k<$tokens$\mbox{-}$count$\left(p_1,p_2,\cdots,p_n\right)$ &
	\begin{tabular}[l]{@{}l@{}}1. the net is 1-safe\\ 2. $k>n$\end{tabular} &
	$false$ \\ 
	\multicolumn{1}{l}{$k\le$tokens$\mbox{-}$count$\left(p_1,p_2,\cdots,p_n\right)$} &
	\begin{tabular}[l]{@{}l@{}}1. the net is a NUPN\\ 2. $k>|\{unit(p_i)\mid1\le i\le n\}|$\end{tabular} &
	$false$ \\
\bottomrule
\end{tabular}}
\end{table}

EnPAC then evaluates atomic propositions as TABLE \ref{tab_APevaluation}, whether they are permanent $true$ or $false$ ($true$ or $false$ under any reachable marking) according to the structural properties of the Petri net. 
For example, if an atomic proposition is the form of $0\le tokens\mbox{-}count(p_1,p_2,\cdots, p_n)$, then it is permanent $true$. The reason is that the number of tokens of any place under any marking is greater than or equal to zero, let alone the sum of tokens of the $n$ places.
Another example is that if a net is 1-safe and an atomic proposition is the form of $tokens\mbox{-}count\left(p_1,p_2,\cdots,p_n\right)\le k$, then it is permanent $true$ when $k\geq n$. Because the upper bound of any place in the 1-safe net is $1$, the token sum of $n$ places must be less than $n$. Moreover, since $k\geq n$, the atomic proposition is permanent $true$. 
If there are any permanent $true$ or $false$ atomic propositions in the formula, then the corresponding atomic propositions can be replaced with $true$ or $false$, and the LTL formula can be iteratively simplified.
Moreover, we propose some further simplification in TABLE \ref{tab_further_simplification}.

\begin{table}[h]\centering
\caption{LTL Formula Further simplification rules}\label{tab_further_simplification}
\begin{tabular}{c|c}
\toprule
$G(true)\equiv true$ & $G(true)\equiv true$ \\ 
$F(true)\equiv true$ & $F(false)\equiv false$ \\ 
$X(true)\equiv true$ & $X(false)\equiv false$ \\
$\varphi\ U\ (true)\equiv true$ & $\varphi\ U\ (false)\equiv false$ \\ 
$true\vee\varphi\equiv true$ & $false\land\varphi\equiv false$ \\
\bottomrule
\end{tabular}
\end{table}

\subsection{Büchi automaton}\label{app_ba}

We implemented the \textit{fast LTL2BA algorithm} \cite{Gastin2001Fast} in EnPAC for translating the simplified LTL formula to a Büchi automaton. We also use the simplification rules in \cite{Gastin2001Fast} to simplify the Büchi automaton. EnPAC implements the on-the-fly framework. One significant feature of on-the-fly is that once a counterexample is found, it stops state-space exploration. So the sooner the counterexample is found, the fewer states are generated. To generate fewer states, we add heuristic information into all Büchi automaton states to guide exploration to find counterexamples. We detail the heuristic Büchi automaton in Section \ref{subsec_heuristicBA}.

\subsection{Product automaton}\label{app_pa}

Product automation generation and counterexample search are also integrated into on-the-fly. 
We implement the \textit{more efficient on-the-fly algorithm} \cite{Geldenhuys2005More} to search counterexample. 

Meanwhile, we adopt the idea of bounded model checking (BMC). BMC \cite{Clarke2001Bounded} unrolls the finite state machine (FSM) for a fixed number of steps $k$ and checks whether a property violation can occur in $k$ or fewer steps. The process can be repeated with larger and larger values of $k$ until all possible violations have been ruled out. 
We leverage the core idea of bounded model checking, i.e., a depth bound $k$. 

EnPAC firstly depth-first checks counterexamples within $k$ steps. Suppose there are no counterexamples within the bound $k$. EnPAC expands the bound. Then EnPAC starts the search from scratch again until a counterexample is detected or the entire state space is generated. 
The advantage is that it can find short counterexamples quickly.

\section{Complete Experimental Result}\label{app_results}

\begin{table*}[t]\centering
\caption{Results of ORI and DYN}\label{tab_DYN}
\begin{tabular}{cc|cccc|cc}
\toprule
\multicolumn{2}{c|}{\textbf{Instances}} & \multicolumn{2}{c}{\textbf{Time (s)}}   & \multicolumn{2}{c}{\textbf{Memory (MB)}}  & \multicolumn{2}{|c}{\textbf{Comparison}} \\ \midrule
Name & Formula & \textbf{$T_{ORI}$} & \textbf{$T_{DYN}$} & \textbf{$M_{ORI}$} & \textbf{$M_{DYN}$} & \textbf{$\nabla T_1$} & \textbf{$\nabla M_1$} \\ \midrule

\multirow{2}{*}{AutoFlight-PT-04a} &
  1 &  101.882 &  95.125 &  3428.82 &  816.023 &  1.07 &  4.20 \\
& 2 & 158.778                    & 147.119 & 3632.53                      & 1019.41   & 1.08     & 3.56     \\
\multirow{2}{*}{BART-PT-020}              & 1 & 0.0206                     & 0.00575 & 808.09                       & 669.086   & 3.58     & 1.21     \\
& 2 & 6.562                      & 1.142   & 7295.99                      & 686.547   & 5.75     & 10.63    \\
\multirow{2}{*}{BART-PT-030}              & 1 & 0.0383                     & 0.00468 & 858.086                      & 651.281   & 8.18     & 1.32     \\
& 2 & 0.0429                     & 0.00771 & 960.781                      & 674.102   & 5.56     & 1.43     \\
\multirow{2}{*}{BART-PT-040}              & 1 & 6.323                      & 5.143   & 2166.43                      & 697.957   & 1.23     & 3.10     \\
& 2 & 0.0381                     & 0.00305 & 1022.39                      & 675.137   & 12.49    & 1.51     \\
\multirow{2}{*}{CircadianClock-PT-000100} & 1 & 2.529                      & 1.731   & 753.027                      & 647.434   & 1.46     & 1.16     \\
& 2 & 0.562                      & 0.529   & 681.512                      & 639.156   & 1.06     & 1.07     \\
\multirow{2}{*}{CircadianClock-PT-001000} &
  1 &
  \textit{\textgreater{}300} &
  123.802 &  3203.16 &  961.637 &  $+\infty$ &  3.33 \\
& 2 & 5.449                      & 2.355   & 1065.21                      & 722.754   & 2.31     & 1.47     \\
\multirow{2}{*}{Dekker-PT-020}            & 1 & 0.351                      & 0.313   & 757.039                      & 641.145   & 1.12     & 1.18     \\
& 2 & 5.493                      & 1.807   & 3092.35                      & 681.301   & 3.04     & 4.54     \\
\multirow{2}{*}{RefineWMG-PT-007007}      & 1 & 1.560                      & 1.130   & 913.371                      & 684.43    & 1.38     & 1.33     \\
& 2 & 115.066                    & 61.484  & 13760.2                      & 2504.87   & 1.87     & 5.49     \\ \midrule
\multicolumn{2}{c|}{Average} & 44.04 (Timeout as 300) & 27.61 & 2774.94 & 835.77 & 3.41 (excluding $+\infty$) & 2.91 \\
\bottomrule
\end{tabular}
\end{table*}

\begin{table*}[t]\centering
\caption{Results of ORI and DRW}\label{tab_DRW}
\begin{tabular}{ccc|cccc|cc}
\toprule
\multicolumn{3}{c|}{\textbf{Instances}}   & \multicolumn{2}{c}{\textbf{Time(s)}}   & \multicolumn{2}{c}{\textbf{Memory (MB)}}        & \multicolumn{2}{|c}{\textbf{Comparison}}   \\ \midrule
Encoding Type & Name  & Formula & $T_{ORI}$ & $T_{DRW}$ & $M_{ORI}$ & $M_{DRW}$ & \textbf{$\nabla T_2$} & \textbf{$\nabla M_2$} \\ \midrule
\multirow{16}{*}{1-safe encoding} &
  \multirow{2}{*}{\begin{tabular}[c]{@{}c@{}}SmallOperatingSystem-\\      PT-MT0064DC0016\end{tabular}} &
  1 &  0.00413 &  0.000793 &  633.105 &  635.074 &  5.21 &  1.00 \\
 &   & 2 & 0.00545  & 0.00148    & 652.082 & 654.641 & 3.68    & 1.00 \\
 &
  \multirow{2}{*}{\begin{tabular}[c]{@{}c@{}}SmallOperatingSystem-\\      PT-MT0128DC0064\end{tabular}} &
  1 &  0.00131 &  0.00106 &  649.977 &  650.297 &  1.24 &  1.00 \\
 &   & 2 & 0.0108   & 0.00218    & 649.977 & 650.297 & 4.95    & 1.00 \\
 & 
 \multirow{2}{*}{\begin{tabular}[c]{@{}c@{}}SmallOperatingSystem-\\      PT-MT4096DC1024\end{tabular}} &
  1 &  0.0288 &  0.00522 &  651.000 &  672.086 &  5.52 &  0.97 \\
 &                                              & 2 & 0.0281   & 0.00439    & 650.906 & 672.016 & 6.40    & 0.97 \\
 & \multirow{2}{*}{SquareGrid-PT-020102}        & 1 & 0.212    & 0.00988    & 633.887 & 635.828 & 21.46   & 1.00 \\
 &                                              & 2 & 0.0126   & 0.000905   & 642.289 & 651.512 & 13.92   & 0.99 \\
 & \multirow{2}{*}{SafeBus-PT-03}               & 1 & 1.505    & 0.0305     & 634.383 & 634.234 & 49.34   & 1.00 \\
 &                                              & 2 & 7.738    & 0.149      & 634.773 & 634.648 & 51.93   & 1.00 \\
 & \multirow{2}{*}{SafeBus-PT-06}               & 1 & 7.210    & 0.0317     & 645.199 & 676.488 & 227.44  & 0.95 \\
 &                                              & 2 & 70.045   & 0.599      & 645.543 & 676.883 & 116.94  & 0.95 \\
 & \multirow{2}{*}{SafeBus-PT-10}               & 1 & 5.841    & 0.00991    & 638.418 & 638.324 & 589.40  & 1.00 \\
 &                                              & 2 & 10.301   & 0.0151     & 650.309 & 721.465 & 682.19  & 0.90 \\
 & \multirow{2}{*}{SafeBus-PT-15}               & 1 & 28.542   & 0.0232     & 642.215 & 642.121 & 1230.26 & 1.00 \\
 &                                              & 2 & 51.707   & 0.0844     & 643.047 & 642.949 & 612.64  & 1.00 \\
\midrule
\multirow{20}{*}{NUPN encoding} &
  \multirow{2}{*}{ARMCacheCoherence-PT-none} &
  1 &   0.908 &   0.0264 &   648.438 &   657.289 &   34.39 &  0.99 \\
 &                                              & 2 & 17.959   & 0.505      & 648.617 & 655.984 & 35.56   & 0.99 \\
 & \multirow{2}{*}{AirplaneLD-PT-0010}          & 1 & 12.308   & 0.470      & 640.578 & 638.812 & 26.19   & 1.00 \\
 &                                              & 2 & 1.276    & 0.0207     & 637.832 & 636.051 & 61.64   & 1.00 \\
 & \multirow{2}{*}{AirplaneLD-PT-0020}          & 1 & 5.372    & 0.0981     & 638.441 & 636.66  & 54.76   & 1.00 \\
 &                                              & 2 & 83.612   & 0.845      & 655.656 & 655.543 & 98.95   & 1.00 \\
 & \multirow{2}{*}{AutoFlight-PT-01b}           & 1 & 0.770    & 0.0254     & 649.395 & 659.422 & 30.31   & 0.98 \\
 &                                              & 2 & 18.439   & 0.684      & 654.566 & 660.277 & 26.96   & 0.99 \\
 & \multirow{2}{*}{AutoFlight-PT-04a}           & 1 & 0.198    & 0.00556    & 648.602 & 652.301 & 35.61   & 0.99 \\
 &                                              & 2 & 0.126    & 0.00384    & 648.602 & 652.301 & 32.81   & 0.99 \\
 & \multirow{2}{*}{CloudDeployment-PT-3b}       & 1 & 81.724   & 0.927      & 641.43  & 641.348 & 88.16   & 1.00 \\
 &                                              & 2 & 83.314   & 0.932      & 642.609 & 649.879 & 89.39   & 0.99 \\
 & \multirow{2}{*}{CloudDeployment-PT-4a}       & 1 & 1.948    & 0.0751     & 636.82  & 634.652 & 25.94   & 1.00 \\
 &                                              & 2 & 58.580   & 2.430      & 644.137 & 641.992 & 24.11   & 1.00 \\
 & \multirow{2}{*}{CloudReconfiguration-PT-306} & 1 & 0.303    & 0.0117     & 641.359 & 639.203 & 25.90   & 1.00 \\
 &                                              & 2 & 0.0298   & 0.000768   & 640.781 & 638.793 & 38.80   & 1.00 \\
 & \multirow{2}{*}{CloudReconfiguration-PT-308} & 1 & 29.818   & 0.582      & 640.176 & 639.977 & 51.23   & 1.00 \\
 &                                              & 2 & 11.289   & 0.220      & 642.281 & 648.094 & 51.31   & 0.99 \\
 & \multirow{2}{*}{DES-PT-00a}                  & 1 & 98.469   & 1.850      & 657.18  & 676.637 & 53.23   & 0.97 \\
 &                                              & 2 & 3.897    & 0.0572     & 637.656 & 656.941 & 68.13   & 0.97 \\
\midrule
\multirow{10}{*}{P-invariant encoding} &
  \multirow{2}{*}{NQueens-PT-08} &
  1 &  4.406 &   0.268 &   634.137 &   636.227 &   16.44 &   1.00 \\
 &                                              & 2 & 114.298  & 6.897      & 643.41  & 645.523 & 16.57   & 1.00 \\
 & \multirow{2}{*}{NQueens-PT-10}               & 1 & 0.535    & 0.0289     & 636.879 & 640.25  & 18.51   & 0.99 \\
 &                                              & 2 & 0.169    & 0.00738    & 634.195 & 633.934 & 22.90   & 1.00 \\
 & \multirow{2}{*}{Peterson-PT-3}               & 1 & 0.000711 & 0.000199   & 653.758 & 651.277 & 3.57    & 1.00 \\
 &                                              & 2 & 0.000733 & 0.000180   & 654.012 & 651.562 & 4.07    & 1.00 \\
 & \multirow{2}{*}{Philosophers-PT-000200}      & 1 & 9.79662  & 0.00223398 & 642.195 & 684.777 & 4385.28 & 0.94 \\
 &                                              & 2 & 9.527    & 0.0453     & 678.965 & 704.766 & 210.31  & 0.96 \\
 & \multirow{2}{*}{Echo-PT-d04r03}              & 1 & 0.519    & 0.000824   & 645.988 & 661.977 & 629.85  & 0.98 \\
 &                                              & 2 & 30.464   & 0.0216     & 646.098 & 658.707 & 1410.37 & 0.98 \\ \midrule
\multicolumn{3}{c|}{Average} & 18.77 & 0.39 & 644.82 & 652.83 & 245.52 & 0.99 \\
\bottomrule
\end{tabular}
\end{table*}

\begin{table*}[t]\centering\caption{Results of ORI and HBA}\label{tab_HBA}
\begin{tabular}{cc|cccccc|ccc}
\toprule
\multicolumn{2}{c|}{\textbf{Instances}} &
		\multicolumn{2}{c}{\textbf{Number of states}} &
		\multicolumn{2}{c}{\textbf{Times (s)}} &
		\multicolumn{2}{c}{\textbf{Memory (MB)}} & \multicolumn{3}{|c}{\textbf{Comparison}}\\ \midrule
Name  & Formula & $N_{ORI}$ &
		$N_{HBA}$ &
		$T_{ORI}$ &
		$T_{HBA}$ &
		$M_{ORI}$ &
		$M_{HBA}$ & \textbf{$\nabla N$} &
  \textbf{$\nabla T_3$} &
  \textbf{$\nabla M_3$} \\ \midrule
ASLink-PT      & 1 & 113532   & 113306 & 0.453                      & 0.458   & 678.242 & 678.383 & 1.00       & 0.99      & 1.00  \\
-PT-02a        & 2 & 321725   & 238268 & 1.684                      & 1.526   & 835.180 & 804.688 & 1.35       & 1.10      & 1.04  \\
Angiogenesis   & 1 & 1045     & 990    & 3.64E-3                    & 3.12E-3 & 633.891 & 633.844 & 1.06       & 1.17      & 1.00  \\
-PT-20         & 2 & 826108   & 142831 & 7.671                      & 1.088   & 1220.34 & 733.934 & 5.78       & 7.05      & 1.66  \\
AutoFlight     & 1 & 265583   & 265468 & 1.418                      & 1.451   & 793.445 & 794.117 & 1.00       & 0.98      & 1.00  \\
-PT-01b        & 2 & 6222621  & 68     & \textit{\textgreater{}300} & 7.71E-4 & 3650.78 & 702.203 & 91509.13   & $+\infty$ & 5.20  \\
AutoFlight     & 1 & 404      & 15     & 3.18E-3                    & 1.82E-4 & 635.934 & 633.758 & 26.93      & 17.47     & 1.00  \\
-PT-03a        & 2 & 621      & 176    & 4.97E-3                    & 1.93E-3 & 636     & 633.852 & 3.53       & 2.58      & 1.00  \\
AutoFlight     & 1 & 666      & 154    & 3.76E-3                    & 8.95E-4 & 706.562 & 707.035 & 4.32       & 4.20      & 1.00  \\
-PT-03b        & 2 & 15354260 & 4427   & \textit{\textgreater{}300} & 7.59E-3 & 7202.09 & 710.859 & 3468.32    & $+\infty$ & 10.13 \\
CSRepetitions  & 1 & 1888     & 41     & 0.0361                     & 9.07E-4 & 637.469 & 636.832 & 46.05      & 39.80     & 1.00  \\
-PT-02         & 2 & 16       & 13     & 9.86E-4                    & 1.01E-4 & 637.797 & 637.148 & 1.23       & 9.76      & 1.00  \\
CSRepetitions  & 1 & 288      & 226    & 1.52E-3                    & 1.53E-3 & 638.512 & 636.234 & 1.27       & 0.99      & 1.00  \\
-PT-03         & 2 & 498814   & 267380 & 2.855                      & 1.328   & 864.184 & 755.465 & 1.87       & 2.15      & 1.14  \\
CircadianClock & 1 & 5480306  & 2      & \textit{\textgreater{}300} & 1.96E-4 & 3228.06 & 634.93  & 2740153.00 & $+\infty$ & 5.08  \\
-PT-001000     & 2 & 3003     & 1001   & 0.0231                     & 0.0104  & 681.77  & 685.766 & 3.00       & 2.22      & 0.99  \\
CircularTrains & 1 & 1.28E+7  & 243    & \textit{\textgreater{}300} & 7.70E-4 & 6282.94 & 751.906 & 52674.90   & $+\infty$ & 8.36  \\
-PT-048        & 2 & 337      & 241    & 7.60E-4                    & 9.21E-4 & 773.793 & 768.605 & 1.40       & 0.83      & 1.01  \\
RefineWMG      & 1 & 1303     & 7      & 0.00187                    & 1.84E-4 & 731.551 & 738.414 & 186.14     & 10.16     & 0.99  \\
-PT-005005     & 2 & 118480   & 116541 & 0.165                      & 0.164   & 788.566 & 794.352 & 1.02       & 1.01      & 0.99  \\ \midrule
\multicolumn{2}{c|}{\multirow{2}{*}{Average}} & \multirow{2}{*}{2100550} & \multirow{2}{*}{57569.9} & 60.72 & \multirow{2}{*}{0.30} & \multirow{2}{*}{1612.86} & \multirow{2}{*}{703.62} & \multirow{2}{*}{144404.62} & 6.4 & \multirow{2}{*}{2.28} \\
\multicolumn{2}{c|}{} & & & (Timeout as 300) & & & & & (excluding $+\infty$) & \\
\bottomrule
\end{tabular}
\end{table*}

\end{document}